\documentclass[%
reprint,
showpacs,preprintnumbers,
amsmath,amssymb,
aps,
prd,
floatfix,
]{revtex4-1}

\usepackage{graphicx,wrapfig,bm,mathrsfs}

\newcommand{\ee}[1]{\begin{align}#1\end{align}}

\begin{document}

\title{Shape Is Destiny II: Configurational Information Approach to Instantons and False Vacuum Decay in D-dimensional Spacetime}
\author{Marcelo Gleiser}
\email{Marcelo.Gleiser@dartmouth.edu}
\affiliation{Department of Physics and Astronomy\\ Dartmouth College,Hanover, NH 03755, USA}

\author{Damian Sowinski}
\email{Damian.Sowinski@dartmouth.edu}
\affiliation{Department of Physics and Astronomy\\ Dartmouth College,Hanover, NH 03755, USA}

\date{\today}

\begin{abstract}
We extend the use of Configurational Information Measures (CIMs) to instantons and vacuum decay in arbitrary spatial dimensions. We find that both the complexity and the information content in the shape of instanton solutions have distinct regions of behavior in parameter space, discriminating between qualitative thin and thick wall profiles. For Euclidean spaces of dimension $D$, we show that for $D\geq 6$ instantons undergo a qualitative change of behavior from their lower-dimensional counterparts, indicating that $D=6$ is a critical dimension.
We also find a scaling law relating the different CIMs to the rate of vacuum decay, thus connecting the stability of the vacuum to the informational complexity stored in the shape of the related critical bounce.
\end{abstract}
\maketitle
\tableofcontents
%---------------------------------------------------------------------------------------------SECTION 1
\section{Introduction}
Information theory made its debut in 1948 with Claude Shannon's revolutionary {\it A Mathematical Theory of Communication} \cite{Shannon:1948}.
In it Shannon was able to quantify information, and, in particular, how it can be transferred from source to receiver through a generalized communication channel. 
In so doing, he proved the celebrated noiseless and noisy coding theorems. 
Central to his analysis was entropy: a measure of how much information is hidden in a random process. 
His insights form the foundation of modern compression technology, but the generality of his work has allowed similar techniques to be brought to bear on a wide array of problems \cite{Huffman:1952}.
Its usefulness in explaining the complexity of neural firing patterns in vertebrate retina, Supreme Court voting patterns, and even natural flocks of birds only begins to address possible applications in very different research fields \cite{Tkacik:2015,Bialek:2015,Bialek:2012}.
Generalizations of Shannon entropy abound in the literature, the most familiar probably being those of Renyi and Tsallis \cite{Renyi:1960,Tsallis:1988}.
Permutation entropy has been developed to study time series, finding uses in quantifying the complexity of plasma turbulence \cite{Bandt:2002,Weck:2015}.
Transfer entropy captures the dynamics of information flow and has found uses in chemical networks within cells \cite{Schreiber:2000,Davies:2013,Davies:2015}.
Parallel to these initiatives, recent work has brought Shannon's information narrative into the context of field theory. This approach is the focus of this paper.

Configurational Information Measures (CIMs) have grown out of the desire to quantify the informational content and complexity contained in the shape of physical structures naturally occurring in field theories.
Constructed in momentum space, configurational entropy (CE) and configurational complexity (CC), and their continuum differential variants, DCE and DCC, are at the core of CIMs \cite{Gleiser:2012}.
CE quantifies the number of bits necessary to construct a field configuration out of wave modes, while CC quantifies the complexity of that construction.
There has been quite a bit of confusion within the literature between these two, which we hope to remedy in the present work with more precise definitions and renamings.

CC was originally proposed by Gleiser and Stamatopolous as a diagnostic tool to compare exact and approximate soliton-like solutions by breaking the degeneracy of equal energy ansatze in models with a single scalar field \cite{Gleiser:2012}, and was soon extended to models with two fields \cite{Correa:2014}, as well as applied to predicting atomic decay rates in hydrogenic atoms \cite{Gleiser:2017a}.

At the same time, the measure's ability to serve as a pattern discriminator has been used to detect the formation of localized structures during inflationary preheating \cite{Gleiser:2014}, as well as to predict the existence and lifetimes of scalar field configurations known as oscillons \cite{Sowinski:2018b}.
CC also serves as a tool for parameter estimation, putting bounds on Lorentz and CPT violation, Abelian string-vortices, $f(R)$ models, and Gauss-Bonnet braneworlds \cite{Correa:2015,Correa:2016,Correa:2015b,Correa:2016b,Correa:2016c}.
The discovery by Gleiser and Sowinski \cite{Sowinski:2013} that CC correlates with the Chandrasekhar limit of white dwarfs hinted at its ability to diagnose instability.
This ability was corroborated in other settings, including $Q$-balls \cite{Sowinski:2013} and, in astrophysics, for neutron and boson stars \cite{Gleiser:2015}, as well as AdS black holes \cite{Braga:2017}.

Meanwhile, CE was shown to play a key role in understanding the information dynamics of phase transitions \cite{Sowinski:2015}.
Quantifying the flow of information between scales, CE shed light on criticality in $2d$ Landau-Ginzburg models \cite{Sowinski:2015}, leading to the concept of information turbulence near criticality \cite{Sowinski:2017a}.
The same methodology identified a phase transition in thick brane models of gravity \cite{Cruz:2017}.
An introduction to CE's relationship to other information measures, and its interpretation as an epistemic tool for the study of structural and dynamical information can be found in \cite{Sowinski:2017b,Sowinski:2018a}.

In this paper, we will add to the phenomenology of CIMs by elucidating the informational narrative of false-vacuum decay via instanton tunneling. 
Instantons are the key ingredient in understanding vacuum decay in field theory, and have been thoroughly examined since they were first introduced by Belavin, Polyakov, Schwartz, and Tyupkin in 1975  \cite{Belavin:1975}. 
Soon after, Coleman showed how, in a semi-classical approach, spherically-symmetric instantons dominate the path integral describing the matrix element connecting distinct potential minima at zero temperature, calculating both the classical and quantum contributions to the false-vacuum decay rate \cite{Coleman:1977a,Coleman:1977b}.

The idea that the vacuum could spontaneously decay into a lower energy state, or that it might have done so in the early universe, suggests that instantons are relevant in a cosmological context \cite{Turner:1982}. 
Indeed, this was the original motivation of Guth's first model of inflation \cite{Guth:1981}, and many that followed. 
Instantons, in particular  of the Fubini-Lipatov type \cite{Fubini1976}, may also play a role in the stability of our own vacuum: within the Standard Model, the value of the Higgs mass implies that the universe is in a metastable vacuum, deeming them important for understanding the fate of our universe \cite{Degrassi:2012,DILUZIO2016150}.
Fortunately, within our current understanding of the Standard Model, any such doomsday scenarios are in the incomprehensibly distant future.
Calculations suggest no transition for $10^{59}$ years at $95\%$ confidence level \cite{Andreassen:2018}.
There exists a sizable literature reviewing these objects, and the decay of the Standard Model vacuum \cite{Kolb:1990,Vainshtein:1982,Coleman:1988,Degrassi:2012,Buttazzo:2013,Paranjape:2017}.

In this paper, we examine how the informational content stored in instanton configurations relate to their physical properties in an arbitrary number of spatial dimensions. 
We will show that both measures of configurational information can shed light into fundamental aspects of instantons and false vacuum decay, including the longevity of the false vacuum. In particular, we obtain a scaling law relating the two which is independent of spacetime dimensionality (for $D\leq 5$, as we will see).

This paper is organized as follows:
Section \ref{s2} reviews the basics of information theory, and clarifies some previous confusion in the current literature about the different configurational information measures (CIMs) and their definitions.
We formalize the difference between entropy and complexity, and elucidate the extension of these measures to the continuum.
Section \ref{s3} introduces the representative scalar field model used in this paper and the conventions we adhere to. 
We also review the construction of an instanton in relation to vacuum decay, and describe the numerical scheme we use in our investigation.
Section \ref{s4} describes our main results, covering both the physical and informational narratives to vacuum decay, and describes how we connect the two through a series of scaling laws.
Section \ref{s5} summarizes our results and discusses future directions. 
In the Appendix we present the calculation of the instanton action in $D$ Euclidean dimensions in the thin-wall limit.
%---------------------------------------------------------------------------------------------SECTION 2
\section{Configurational Information Measures}\label{s2}
\subsection{Entropy, Information, and Complexity}
Central to proving the noisy and noiseless coding theorems in information theory is a quantity called Shannon entropy:
\ee{\label{Shannon Entropy}
S = -\sum_{a\in \mathcal A}p_a\log p_a
.} 
Here, $\mathcal A$ is an alphabet of symbols, $X$ is a random symbol drawn from that alphabet, and $p_a=p(X=a|\mathcal L)$ is the probability of that symbol given a language $\mathcal L$. 
The base of the logarithm is a matter of preference: computer scientists tend to use base $2$ and measure entropy in {\it bits}, while information theorists and physicists prefer base $e$ and measure entropy in {\it nats}.
Note that $S$ is bounded from above by $\log|\mathcal A|$, the case when each symbol is equiprobable, and from below by $0$, when a single symbol is certain.
The latter lends itself to the $\mathscr S$-interpretation of entropy as {\it surprise}: an event that is certain is not surprising.
Another viewpoint is the $\mathscr Q$-interpretation, in bits, where entropy represents the average number of yes-no questions one must ask to determine a random letter.
Random here means the letter is drawn from some distribution, $p$, the analogue to a language.
A good example is English, where one imagines a letter is drawn from the corpus of English literature.
The entropy is a little over 4 bits, meaning that it will take on average a little over 4 guesses to determine the identity of the letter \cite{Sowinski:English,Shannon:1951}.

Shannon's theorems prove that entropy provides a lower bound on how compressible an alphabet is in a given language; it can thus be interpreted as the amount of information needed to encode a random phenomenon. 

The information contained in a particular occurrence of a letter of the alphabet is $I(a) = -\log p_a$.
Improbable letters carry a lot of information.
If this seems a bit abstract, try to place it in the context of a game of Scrabble.
Revealing an $e$ doesn't pin down the possible words you could make since there are a great many possibilities for incorporating that $e$ into your next move. 
Revealing an $x$, however, is a completely different story.
Future moves are constrained much more by this event.
It is this constraining of possibility that the notion of information attempts to capture. 

Entropy and information are related in that the former is the expected value of the latter:
\ee{\label{Information}
\langle I\rangle = \sum_{a\in \mathcal A}p_a I(a)=-\sum_{a\in\mathcal A}p_a\log p_a = S
.}
The information content of a letter turns out to be the optimal number of bits needed to encode it in order to optimize the transmission of information through a communication channel, thus saturating the bound revealed by entropy.

\begin{figure*}[htp]
\centering \includegraphics[width=\textwidth]{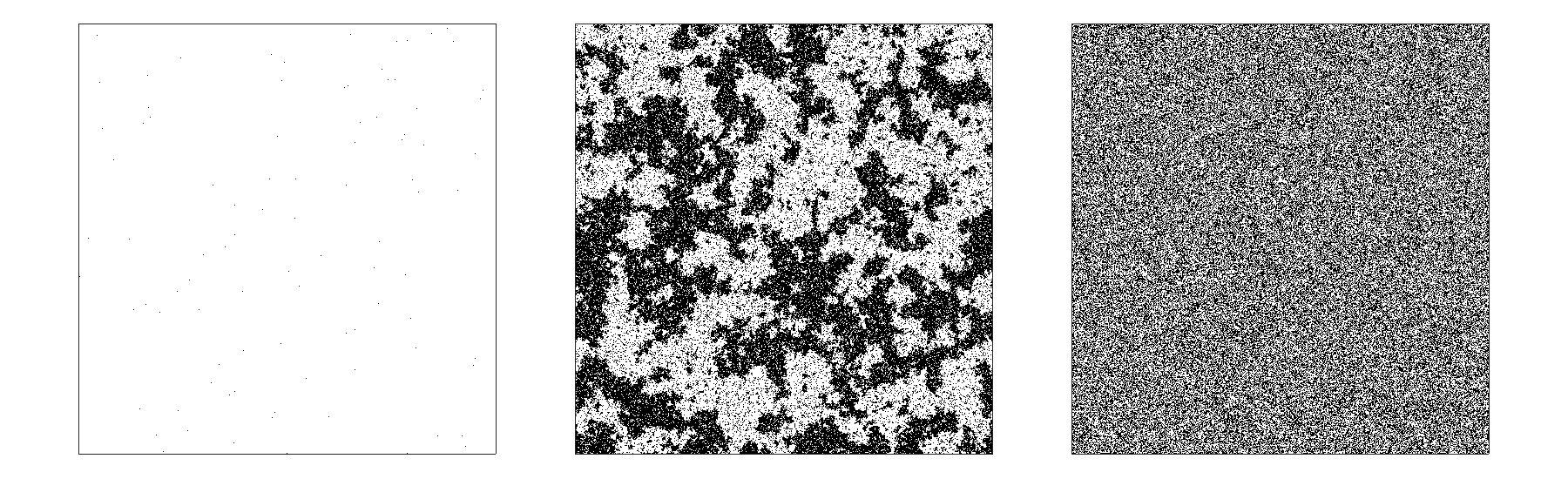}
\caption{The Ising model at different temperatures as an example of complexity. To the left the model is at a temperature $T\ll T_c$, while to the right $T\gg T_c$. In both cases the description of the system is short, signifying a low complexity. In the central picture the model is close to criticality, $T\approx T_c$, and the scale invariant structure due to long range correlations makes the description of the image longer - this is the hallmark of greater complexity. }
\label{Ising Model}
\end{figure*}

The formal study of complexity had its infancy in a series of papers published by Andrey Kolmogorov in the 1960s \cite{Kolmogorov:1963, Kolmogorov:1965, Kolmogorov:1968}.
Closely related to entropy, complexity is defined as the minimal description of an object in all possible languages. 
Both low entropy processes and high entropy ones (close to the $\log {\cal A}$ limit) have low complexity.
A good picture to keep in mind is of the standard Ising model well below or above the critical temperature, as in Figure \ref{Ising Model}.
In the former case, the system is dominated by a single phase, while in the latter it has the appearance of white noise.
Both these descriptions are short and to the point. 
At criticality, however, long range correlations generate scale invariant structures that are hard to describe in few words.
This is a generic feature of complexity measures: they attain a maximum somewhere between minimal and maximal entropy.

\subsection{The Continuum Limit}
What if we now allow our alphabet to grow so that we have before us a language with an infinitude of letters?
If the alphabet is countably infinite, $|\mathcal A| = \aleph_0$ then the upper bound on the entropy diverges, but otherwise there are no complications in interpretation. 

A direct generalization to the continuum by the introduction of a probability density, $p_a = \rho(a)da$, results in a breakdown of the $\mathscr Q$ and $\mathscr S$ interpretations of entropy: the degrees of freedom in the continuum create a divergence.
To see this, simply rewrite Eq. \ref{Shannon Entropy} as
\ee{\label{Continuum Entropy}
S&=-\lim_{da\rightarrow 0}\sum_{a\in \mathcal A}\rho(a)da\log\rho(a)da\nonumber\\
&=-\int da\rho(a)\log \rho(a)+\lim_{da\rightarrow 0} \log\frac{1}{da}
.}
The second term introduces a logarithmic divergence, compromising the continuum generalization. 
However, this divergence may not be an issue: in physical applications, there are often natural cutoff scales that would take care of it. 
For example, there may be a maximum length scale (minimum momentum scale), or an effective coarse-graining scale.

Indeed, differential entropy (``differential'' is added in the literature to identify continuum measures) simply ignores this infinite shift, and uses the first term in Eq. \ref{Continuum Entropy},
\ee{
\mathcal S = \int\! da \ \rho(a)\log\rho(a)
.}
Though finite, care must be taken in its interpretation.
Since it is not invariant under a change of coordinates, it is relative to a given coordinate system. 
This is clear once one recalls that the probability density transforms as a scalar density under coordinate transformations: when $x\rightarrow\bar x$, the density transforms as $\rho(x)\rightarrow |\frac{\partial \bar x}{\partial x}|\bar\rho(\bar x)$.
Also, it is not positive-definite, making both $\mathscr Q$ and $\mathscr S$ interpretations problematic.

One way around these pitfalls is through the introduction of a reference distribution as in the Kullback-Leibler divergence in information theory \cite{Kullback:1951}.
CIMs, however, follow a different approach, viewing both issues--coordinate-dependence and non-positivity--as features.
Configurational Entropy is indeed dependent on the coordinate system being used to describe a given physical object.
Recalling the language analogy, phonemes requiring one symbol in one alphabet may require two symbols in another:
information is measured relative to a {\it fixed} alphabet. Likewise, even if the shape of a field configuration is invariant, its description in different coordinate systems will entail a different amount of information. This is why we instinctively seek the most symmetric coordinate system in a given physical situation, to minimize the information needed for its description.

A non-positive differential entropy within the context of CIMs is more a statement about the physical properties of the model under study than an issue with the formalism. 
Indeed, negative values of configurational entropy are indicative of some length scale in the problem being ``too large.''
A large spatial configuration will have a small resolution in momentum space, making it hard to distinguish from, e.g., a noisy background. 

\subsection{Configurational Information}\label{CI}
Configurational information measures are concerned with the {\it shape} complexity of objects. 
In the context of field theories, a localized object such as a soliton or a topological defect can be thought of as the solution of a single or of a set of coupled nonlinear PDEs with specific boundary conditions. 
This is true for time-dependent solutions as well, as in the case of oscillons. 
Thus, the specific shape of the object is a direct manifestation of the dynamical properties of the system, its interactions and boundary conditions. 
For this reason, CIMs are conjectured to capture dynamical information.

The shape of a structure can be described through its two-point correlation function.
Fourier transforming this, the power spectrum carries an equivalent description in terms of momentum modes.
CIMs are constructed from the modal description of an object.
Let us first derive the configurational entropy measure (CE) in the discrete case, and then use that to define an analogous configurational complexity (CC).  
We will then move on to their differential (continuous) counterparts.

\subsubsection{CE and CC}
Consider a scalar field configuration, $\varphi(\mathbf r)$, within a finite volume, $V$, and periodic boundary conditions.
The field can be decomposed into a countable sum of Fourier modes
\ee{
\varphi(\mathbf r) = \sum_{\mathbf k}\tilde \varphi_{\mathbf k}e^{i\mathbf k\cdot\mathbf r}
.}
As mentioned above, the two-point correlation function contains information about the shape of the field.
Its Fourier transform, the power spectrum, encapsulates the strength of all the modes that go into generating the configuration, 
\ee{
\frac{1}{V}\int\! d^d\mathbf r' \ \varphi(\mathbf r')\varphi(\mathbf r'+\mathbf r)=\sum_{\mathbf k}|\tilde\varphi_{\mathbf k}|^2e^{i\mathbf k\cdot\mathbf r}
.}
Given observers (human or idealized) that can measure the length scales inherent in the field configuration, the probability that they measure a particular scale is proportional to that scale's power,
\ee{
p_{\mathbf k}=\frac{|\tilde\varphi_{\mathbf k}|^2}{\sum_{\mathbf k'}|\tilde\varphi_{\mathbf k'}|^2}
.}
With this in hand, the configurational entropy (CE) is defined as the Shannon entropy of this distribution
\ee{
S_C[\varphi]=-\sum_{\mathbf k}p_{\mathbf k}\log p_{\mathbf k}
.}
This measure vanishes for a single plane wave.
As the number of plane waves increases, so does the CE.
In the limit that the configuration is localized to such an extent that the power spectrum becomes uniform, the CE approaches its maximal value.
The CE represents the information necessary, by "counting" the number of modes, needed in the construction of a configuration.

In the same way that a codex filled with random letters has a high entropy but low complexity, a configuration that is built of many modes need not be complex. 
For example, a delta function has a uniform modal fraction - its complexity, as measured by the length of description, is very small.
Now consider another configuration, with a certain power at wavelengths at the scale of the volume, half as much power at wavelengths half that size, half as much again at wavelengths a quarter size, and so on. 
The description of this configuration is certainly longer than the first, so it has a larger complexity.
This implies that if the power in non-zero modes is uniform, there is little complexity, whereas if the power is distributed non-uniformly among these modes the complexity increases.
Given a specific power spectrum, the relative contribution of different modes is quantified by the {\it modal fraction}
\ee{
f_{\mathbf k} = \frac{|\varphi_{\mathbf k}|^2}{\max |\varphi_{\mathbf k'}|^2},
}
which satisfies $0\le f_{\mathbf k}\le 1$. 
The normalization with the maximum mode guarantees the positivity of the
configurational complexity (CC), defined as 
\ee{
\mathcal C_C[\varphi] = -\sum_\mathbf k f_\mathbf k\ln f_\mathbf k.
}
This measure vanishes if all the non-zero modes contributing to a configuration have a uniform modal fraction (i.e., carry the same weight).
Uncorrelated noise, for example, has a uniform power spectrum.
This means that it has maximal CE, but vanishing CC. 
A plane wave, on the other hand, has both vanishing CE and CC.
Somewhere in between utter monotony and utter randomness is where CC maximizes, lending weight to the interpretation of CC as a measure of shape complexity.
This justifies our use of ``entropy'' and ``complexity'' for CE and CC, respectively \footnote{In the previous literature on configurational entropy, this distinction wasn't clear. 
As a result, what we now call CC was confusingly called CE before.}.
\subsubsection{DCE and DCC}
We now generalize our previous approach to non-periodic square-integrable functions, taking into account the increase in the number of degrees of freedom in momentum space as the limits of integration (the spatial boundaries where the function is defined) grow larger. 
At spatial infinity we have a continuum in momentum space.

The Fourier transform of the field in $d$ spatial dimensions is
\ee{
\tilde \varphi(\mathbf k) = (2\pi)^{-\frac{d}{2}}\int\!\! d^d\mathbf r \  \varphi(\mathbf r) e^{-i\mathbf k \cdot \mathbf r}
.}
The probability of detecting a scale centered at mode $\mathbf k$ and within a volume $d^d\mathbf k$ is proportional to the power in the mode
\ee{
p(\mathbf k,d^d\mathbf k) &= \frac{|\tilde \varphi(\mathbf k)|^2}{\int \! d^d\mathbf k' \ |\tilde \varphi(\mathbf k')|^2}d^d\mathbf k\nonumber\\
&=\rho(\mathbf k)d^d\mathbf k
,}
where we have introduced the probability density, $\rho$.
The Differential Configurational Entropy (DCE) is computed from this density as 
\ee{
\mathcal S_C[\varphi] = -\int \! d^d\mathbf k \ \rho(\mathbf k)\log \rho(\mathbf k)
.}
We need not worry about negative values of DCE: they are associated with unphysical distributions.

For the Differential Configurational Complexity (DCC), we must first specify the modal fraction, normalized by the maximum mode contribution,
\ee{
f(\mathbf k) = \frac{|\rho(\mathbf k)|^2}{\max_{\mathbf k'} |\rho(\mathbf k')|^2}
.}
As before, the DCC is defined as
\ee{
\mathcal C_C[\varphi] = -\int \! d^d\mathbf k \ f(\mathbf k)\ln f(\mathbf k)
.}
Unlike DCE, this quantity is guaranteed to be positive due to the fact that the modal fraction $f(\mathbf k) \leq 1$.

For completeness, we compute the DCC for the case of spherical symmetry, where some care must be taken.
The hyper-spherical Fourier transform reads
\ee{
\tilde\rho(k) = k^{1-\frac{d}{2}}\int_0^\infty\!\!\! dr \ r^\frac{d}{2}\rho(r)J_{\frac{d}{2}-1}(kr)
,}
where $J_\nu$ are Bessel functions. 
A detector sensitive to scale will measure modes with probability $|\tilde\rho(k)|^2d^d\mathbf k$.
The modal fraction is given by
\ee{
f(k) = \frac{| \tilde\rho( k)|^2}{{\rm max}_{k'}|\tilde\rho(k')|^2}
,}
and the DCC is then
\ee{
{\cal C}_C[\rho] = \frac{2\pi^{d/2}}{\Gamma(\frac{d}{2})}\int_0^\infty dk \ k^{d-1}f(k)\log f(k)
.}
%---------------------------------------------------------------------------------------------SECTION 2

\section{Vacuum Decay and Instantons}\label{s3}
%\subsection{The Model}
We consider a real scalar field $\phi$ in a flat spacetime of dimension $D = d+1$.
We write the Minkowski metric as $\eta_{\mu\nu}={\rm diag} (-,+,+\cdots)$.
The action for our model is
\ee{\label{Action}
S[\phi] =\int_\mathcal M\!\!\! dV\left(- \frac{1}{2}\eta^{\mu\nu}\partial_\mu\phi\partial_\nu\phi-V(\phi)\right)
,}
where $dV$ is the invariant volume element on the spacetime region $\mathcal M$, and $V(\phi)$ is a potential with two minima. 
Quantum corrections, if important, are incorporated in the potential parameters.
Since we will examine both degenerate and non-degenerate vacua, we parameterize our potential with the triplet $(\lambda,\phi_0,\epsilon)$ as
\ee{\label{Potential}
V(\phi) = \frac{\lambda}{4}\phi^2(\phi-\phi_0)^2-\epsilon\frac{\lambda}{2}\phi_0\phi^3
.}
The first term is a $\mathbb Z_2$-symmetric potential with vacua at $\phi = 0$ and $\phi_0$. 
The second introduces an asymmetry, parameterized by the dimensionless quantity $\epsilon$, that shifts the latter of these vacua to a lower energy, thereby making the minimum at $\phi=0$ metastable.

\subsection{Dimensionalization Convention}
The triplet of parameters can be reduced to two by choosing an appropriate scale.
To see this, let $\phi(x) = (\phi_0/v) \bar\phi(\bar x)$ and $r=(v/\sqrt{\lambda\phi_0^2})\bar r$, with $v$ an arbitrary dimensionless parameter.
The action of Eq. \ref{Action} divided by $\lambda^\frac{2-D}{2}\phi_0^{4-D}v^{D-2}$ can be written in dimensionless form as
\ee{\label{Dimensionless Action}
\bar S[\bar\phi] = \int \! d\bar V(-\frac{1}{2}&\bar\eta^{\mu\nu}\bar\partial_\mu\bar\phi\bar\partial_\nu\bar\phi\nonumber\\
-&\frac{1}{4}\bar\phi^2(\bar\phi-v)^2+\frac{1}{2}\epsilon v\bar\phi^3)
.}
Henceforth we drop the bars, and it is understood that we work with the dimensionless variables.
A quick note on the dimensionless parameter $v$, which specifies the spacing between vacua of the potential.
Its value is a matter of choice.
Gleiser \cite{Gleiser:1994,Copeland:1995} and Honda \cite{Honda:2000}, for example, choose $v=2$, and shift the field $\phi\rightarrow\phi-1$.
Kolb and Turner, whom we follow here, choose $v = \sqrt{2}$  \cite{Kolb:1990}.

\subsection{Vacuum Decay}
To find the probability of false vacuum decay one must compute the Wick rotated matrix element between the false vacuum at past infinity and the true vacuum at future infinity, resulting in a Euclidean path integral, 
\ee{
\mathcal M_{F\rightarrow T} = \langle \phi_0,+i\infty|0,-i\infty\rangle=\int \! \mathcal D\phi \ \ e^{-S_E[\phi]/\hbar}
.}
In the semi-classical approximation, one uses the saddle-point approximation to show that the path integral is dominated by a single field configuration which minimizes the action, which we denote as the critical instanton or instanton, $\phi_B$.
Configurations that are sufficiently {\it far} from the instanton are exponentially suppressed and do not contribute significantly to the path integral. 
This assumption allows us to expand the action as
\ee{
S_E[\phi] = S_E[\phi_B] + \frac{1}{2}\frac{\delta^2 S_E}{\delta\phi^2}(\phi-\phi_B)^2+\cdots
}
This reduces the matrix integral to a Gaussian approximation (with $\delta \phi = \phi-\phi_B$)
\ee{
\mathcal M_{F\rightarrow T} \approx e^{-S_E[\phi_B]/\hbar}\int\!\mathcal D \delta\phi \ \ e^{-\frac{1}{2\hbar}\int d^Dx\delta\phi(\square + V''(\phi_B))\delta\phi}
.}
The perturbative solution to this integral is known to order $\hbar$, and represents a quantum correction to the decay rate. It appears as a sub-dominant prefactor multiplying the exponential decay \cite{Coleman:1977a,Coleman:1977b}.
The semiclassical decay rate of the false vacuum is the ratio of the transition $F\rightarrow T$ to $F\rightarrow F$,
\ee{
\Gamma \sim \frac{\mathcal M_{F\rightarrow T}}{\mathcal M_{F\rightarrow F}} \sim e^{-S_E/\hbar}.
}
The instanton $\phi_B$ has $O(D)$ symmetry, being a function solely of $\xi = \sqrt{\tau^2+r^2}$, where $\tau$ is the Wick rotated time, and $r$ is the spatial radius.
We therefore seek solutions of
\ee{\label{Instanton EoM}
\frac{d^2\phi}{d\xi^2}+\frac{D-1}{\xi}\frac{d\phi}{d\xi} &= \phi-\frac{3}{\sqrt{2}}(1+\epsilon)\phi^2+\phi^3
,}
with $\phi'(0) = 0$ to ensure regularity at the origin, and $\phi(\xi\rightarrow\infty) = 0$ to match the false vacuum at past and future infinity.
Solutions for varying $\epsilon$  are found numerically using the shooting method. 
As is well-known, one can interpret Eq. \ref{Instanton EoM} treating $\xi$ as time. The equation then describes a particle moving in a potential $-V(\phi)$, with a time-dependent friction force.
The boundary condition at infinity demands that the particle end its journey on the hill corresponding to the $\phi = 0$ vacuum. 
Without friction, the particle would have to be released from the classical turning point, but the presence of friction requires its journey to start at a higher value. That higher value is the core of the instanton, $\phi_c$, the field at $\xi=0$.
\subsection{Numerical Implementation}
\begin{figure}
\includegraphics[width = \columnwidth]{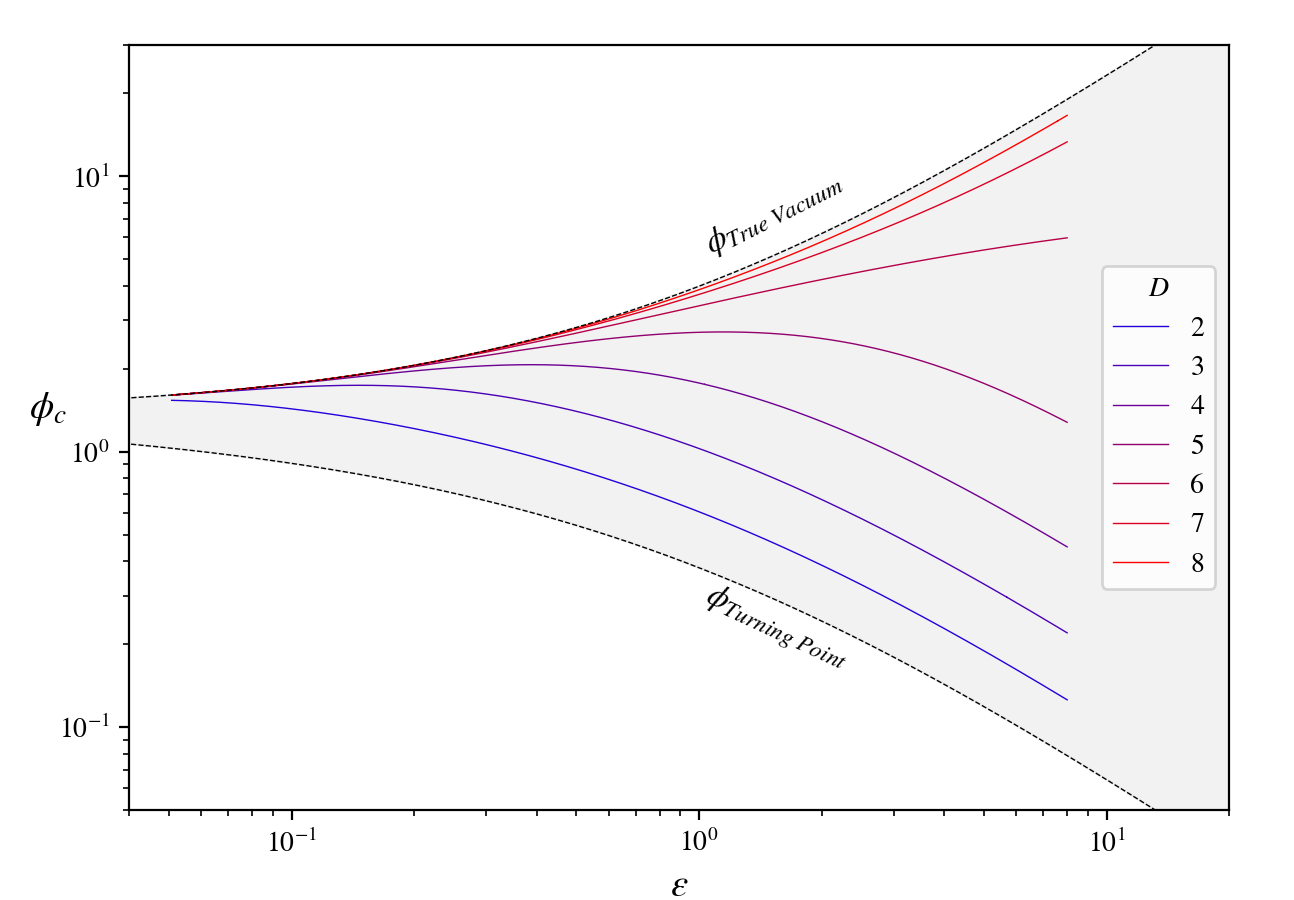}
\caption{The core value of critical instantons lays between the classical turning point and the true vacuum. For small asymmetry the instanton core tracks true vacuum. As asymmetry increases, the core value becomes noticeably smaller than the true vacuum and tracks the turning point. This effect is suppressed at dimensions $D\geq 6$ as the friction term in the EoM increases, resulting in a qualitative change of behavior.}
\label{core_asymmetry}
\end{figure}

Solutions are found by applying a binary search between the classical turning point and the true vacuum to pinpoint the core value of the instanton.
This is done in dimensions $D=d+1\in [2,8]$, and across two orders of magnitude for the asymmetry parameter $\epsilon\in[.05,2]$ using step sizes $\Delta \epsilon = 4.875\times 10^{-4}$ and $\epsilon\in(2,10]$ using $\Delta \epsilon = 1.3\times 10^{-3}$.
Numerical integration is performed using a $4^{th}$-order Runge-Kutta method on the domain $\xi\in [0,80]$ using up to $N=4000$ lattice points.
64-bit floats were used, so for instantons with larger asymmetries numerical convergence was achieved at less than $N=4000$ grid points, each case being treated individually.
Numerical results for the computed core values are plotted in Figure \ref{core_asymmetry}. For $D\geq 6$ we see a qualitative change in behavior, suggesting that $D=6$ is a critical dimension for vacuum decay. (We will say more about this later.)

Fits are done to extract the power law behavior of different regimes of physical parameters.
In order to not be biased towards large magnitudes, our fitting routine is performed in $\log-\log$ space.
To get the error bars, bootstrapping is done with one hundred subsets of data points of random size in said region.
Means and standard deviations on the resulting distribution of measured parameters are then reported. 
%---------------------------------------------------------------------------------------------SECTION 4
\subsection{Instantons} \label{s4}
\begin{figure}[htp]
\includegraphics[width=\columnwidth]{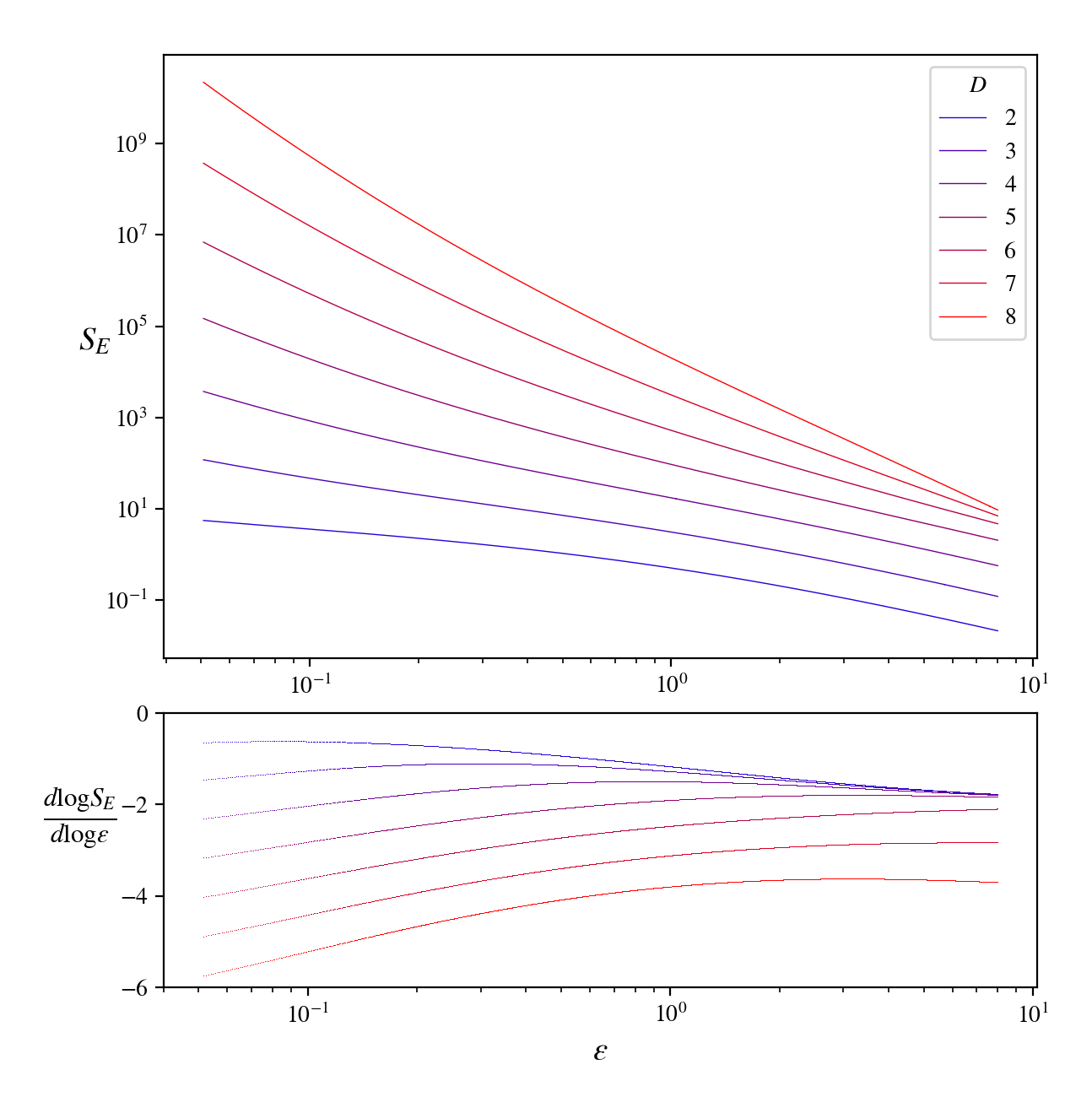}
\caption{The Euclidean action as a function of asymmetry. The lower panel is the power law dependence on asymmetry.}
\label{euclidean_action}
\end{figure}

The stability of the false vacuum is encapsulated in the decay rate -- the greater the rate, the more likely the false vacuum will nucleate a critical instanton that induces a phase transition.
The decay rate is suppressed exponentially by the Euclidean action of the instanton. 
We plot the Euclidean action versus asymmetry in Figure \ref{euclidean_action}.
At small $\epsilon$, the instantons are in the {\it thin wall limit}, and $S_E \sim \epsilon^{-d}$.
The case for $D=4$ is covered in ref. \cite{Kolb:1990}. We derive the general relationship in Appendix \ref{a1}.

For large $\epsilon$ localized configurations are described as {\it thick wall} instantons.
The dependence of the action on asymmetry is close to a power law, with some variation over decades. 
We note that for $D\le 5$ the thick wall power law exponent appears to converge in a dimensionally-independent way. (See bottom panel.) For larger dimensions this behavior changes qualitatively, again indicating the existence of a critical dimension at $D=6$.
We conjecture that this is related to the well-known observation that the volume of a hypersphere of unit radius maximizes at $D=5$, and surface area at $D=7$.
Numerical results indicate that there exists a critical dimension in this range.
\subsection{ Size Scaling and the Critical Dimension }
\begin{figure}[htp]
\includegraphics[width=\columnwidth]{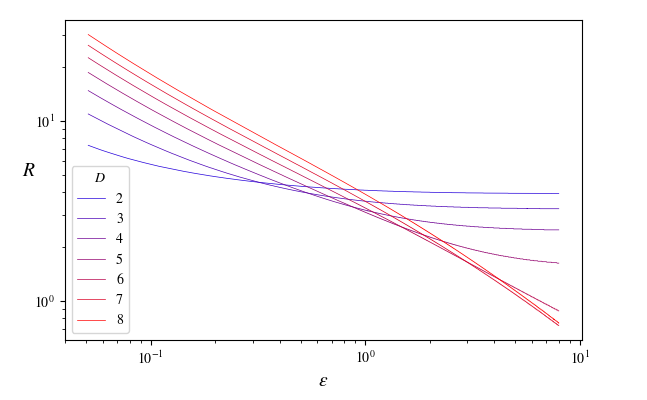}
\caption{The effective radius as a function of asymmetry. Note how low dimensional and high dimensional instanton radii differ in the thick wall limit. }
\label{Radius versus asymmetry}
\end{figure}

The volume of instantons is dominated by negative potential energy, while the wall by positive gradient energy.
The former acts to create an outward pressure on the instanton, while the latter gives the instanton boundary an effective tension which generates an inward compression. 
The Euclidean action is dominated by the contribution coming from the wall, where the magnitude of the derivative of the field attains a maximum.
It is numerically verified that soon after this maximum the field plummets to its false vacuum value.
We use this behavior to define an effective radius for the instanton,
\ee{R = \arg\min_r \left|\frac{\phi'(r)}{\min \phi'}-.01 \right|.\label{Effective Radius}}
We plot the effective radius versus asymmetry in Figure \ref{Radius versus asymmetry}.
We note the same distinction between instantons in $D\leq 5$ and at higher dimensions: 
In the thick-wall limit and for $D\leq 5$, the effective radii become independent of $\epsilon$.
We will focus for now on instantons for $D \leq 5$, planning to present a thorough analysis of effects associated with the critical dimension in future work. 
In this range of dimensions, the asymptotic radius decreases linearly with dimension, reaching a minimum value
\ee{
\lim_{\epsilon\rightarrow\infty}R_{\rm min} = 4.767^{\pm.084}-(D-1)0.777^{\pm.025}.
\label{Rmin}
}
This trend cannot persist since it becomes negative at around $D=7$, once again pointing to the existence of a critical dimension below that. 

\section{The Informational Narrative}
We now investigate how the different information-entropic measures introduced in Section \ref{CI} can inform the physics of false vacuum decay.
In what follows, we will compute the DCC and DCE for various profiles describing instantons. \subsection{Instanton Profiles}
\begin{figure}[htp]
\includegraphics[width=\columnwidth]{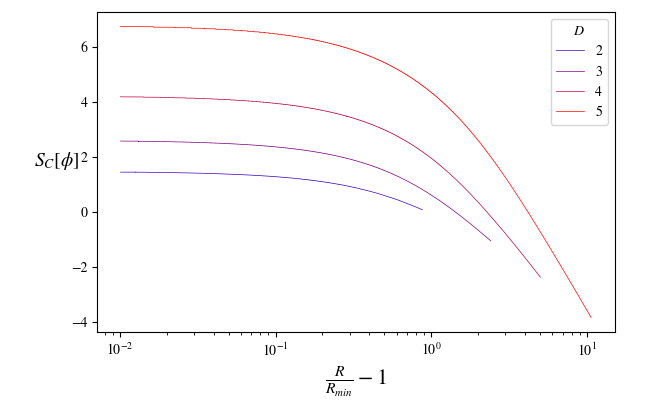}
\caption{The Differential Configurational Entropy as a function of deviation from minimum radius. The plateaus at small deviation indicate an insensitivity to asymmetry in the thick wall regime.}
\label{CE_versus_R}
\end{figure}

Instanton solutions have a rather simple shape. 
In the {\it thin wall} regime, at small asymmetries, the field throughout most of the instanton takes on a constant value at $\phi \simeq \phi_c$.
Within a small range of the radial variable $\xi$, the field quickly transitions into the false vacuum value. 
As $\epsilon \rightarrow 0$, the instanton radius grows without bound. (See the Appendix.)
In configuration space this means that the modal fraction becomes more and more selective.
This decrease in uncertainty about which momentum modes are generating the instanton translates into a decrease in configurational entropy.
On the opposite end, the {\it thick wall} regime, instantons become highly localized in space.
One might imagine that they become progressively smaller with increasing asymmetry, but the DCE indicates otherwise by plateauing. 
This means that the modal fraction is not changing, and hence that the shape of the solutions is insensitive to asymmetry.
This becomes apparent in Figure \ref{CE_versus_R}, where we plot the DCE versus radial deviation from the minimum value of Eq. \ref{Rmin},

\begin{figure}[htp]
\includegraphics[width=\columnwidth]{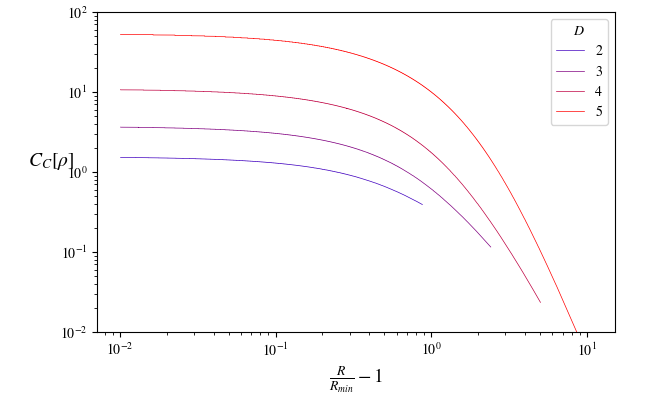}
\caption{The Differential Configurational Complexity as a function of deviation from minimum radius. As the instantons get larger, the modal fraction becomes less complex since a single mode begins to dominate. }
\label{CC_versus_R}
\end{figure}

The DCC paints a similar story, summarized in Figure \ref{CC_versus_R}.
When the modes contributing to the shape of the instanton all contribute nearly equally, its configurational complexity decreases.
Conversely, when different modes making up the configuration carry different weights its configurational complexity grows.

One can understand the plateaus in both DCE and DCC in the thick-wall limit using a simple scaling argument.
Consider the equation of motion Eq. \ref{Instanton EoM}, and perform a rescaling of the field $\phi\rightarrow \phi/\epsilon$.
Since, in the large $\epsilon$ limit, both $\phi_c\ll 1$ and $\epsilon^{-1}\ll1$, one of the quadratic terms and the cubic term are sub-dominant, we can write,
\ee{\frac{d^2\phi}{d{\rho^2}}+\frac{D-1}{\rho}\frac{d\phi}{d\rho} \simeq \phi-\frac{3}{\sqrt{2}}\phi^2.}
We see then that in the thick-wall limit, the effective equation determining the shape of the instantons becomes independent of the asymmetry.
There is still dependence on the spacetime dimensionality, but, for a given dimension $D$, there is only a single instanton at high asymmetry.

\subsection{Relating CIMs to the Vacuum Decay Rate}
\begin{figure}[htp]
\includegraphics[width=\columnwidth]{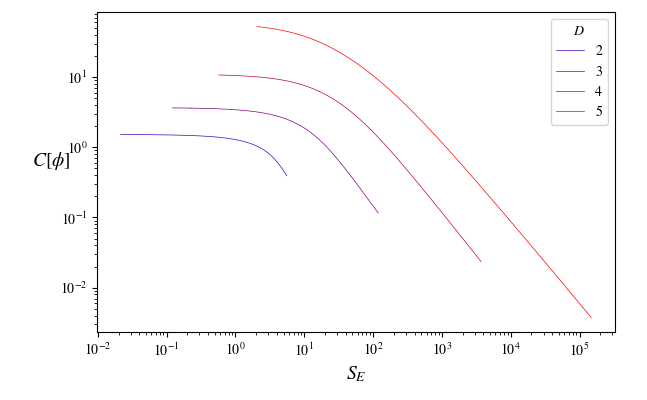}
\caption{The log-log plot of the Differential Configurational Complexity versus Euclidean action for $D\leq 5$. }
\label{CC_fields_vs_SE}
\end{figure}

Given that the vacuum decay rate is dominated by the instanton's Euclidean action, and, in turn, that the Euclidean action is determined by the instanton shape, it must be possible to extract information about the vacuum decay timescale from CIMs. 
In Fig. \ref{CC_fields_vs_SE} we plot the DCC versus Euclidean action for $D\leq 5$.
From the plateaus in the thick-wall limit observed in Figs. \ref{CE_versus_R} and \ref{CC_versus_R}, it is not surprising that we see similar behavior here for small Euclidean actions: there is little change in shape complexity for large $\epsilon$.
More interestingly, one can detect a self-similar behavior across the different dimensions, which we can match to a functional relation approximately modeled by a Lorentzian-like function,
\ee{{\cal C}_C = \frac{{\cal C}_C^\text{max}}{1+b(\frac{S_E}{S_E^\text{min}}-1)^c},
\label{Lorentzian1}}
where $S_E^\text{min}$ is the action of the instanton with minimum radius $R_{\rm min}$ of Eq. \ref{Rmin} (see Figure \ref{Radius versus asymmetry}) and ${\cal C}_C^\text{max}$ is the maximum configurational complexity (see Figure \ref{CC_versus_R}).
For large Euclidean actions typical of the thin-wall limit, there is a clear decrease in DCC, as discussed above. 
We thus expect the exponent $c$ to be positive. 
Below, we obtain analytical estimates for the scaling between DCC and $S_E$ in the thin and thick-wall limits and extract the value of $c$. 

First, note that on dimensional grounds, the DCC should scale with spacetime dimension as $ {\cal C}_C \sim R^{-D}$.
This can be seen, e.g., for a $D$-dimensional Gaussian \cite{Gleiser:2012}, where $C_C = D/2(2\pi/R^2)^{D/2}.$
We verified numerically that this is the case in the thin wall limit, with a best fit of
$ {\cal C}_C \sim (R/R_{\rm min}-1)^{-1.07^{\pm .10}(D-1) + .42^{\pm .09}}$, or ${\cal C}_C \sim R^{-D+3/2}.$

\begin{figure}[htp]
\includegraphics[width=\columnwidth]{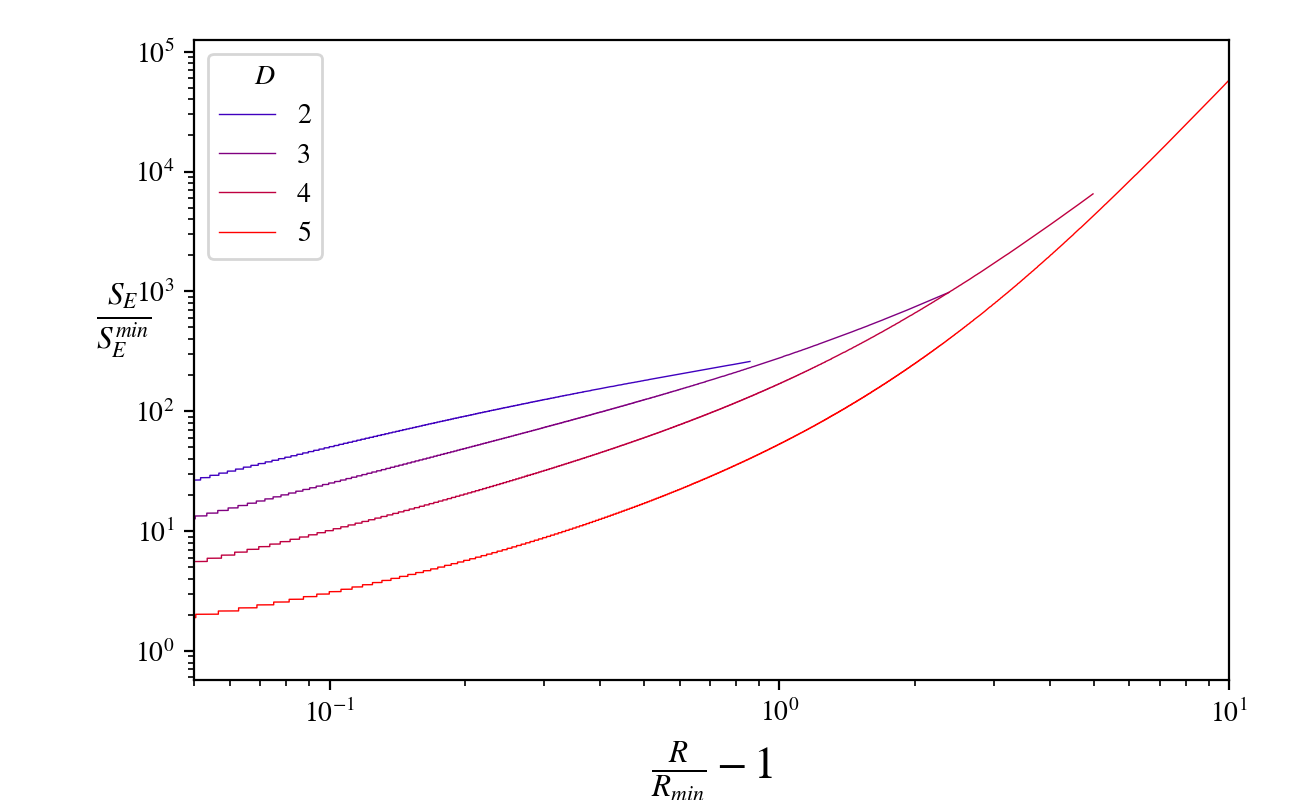}
\caption{The Euclidean bounce action as a function of radius for spacetimes with dimension less than critical.}
\label{SE_versus_R}
\end{figure}

The thin wall limit corresponds simply to $S_E \gg S_E^\text{min}$ ($R\gg R{\rm min}$), or, from the Lorentzian-like fit of Eq. \ref{Lorentzian1}, ${\tilde {\cal C}}_C \sim \tilde S_E^{-c}$, where we introduced ${\tilde {\cal C}}_C \equiv {\cal C}_C/{\cal C}_C^{\text max}$, and $\tilde S_E \equiv S_E/S_E^{\text min}$.
In Fig. \ref{SE_versus_R}, we plot the Euclidean action as a function of the effective radius as defined in Eq. \ref{Effective Radius}. 
For large $R$, we note a natural trend which, can be fitted as
$S_E \sim R^{(D-3/2)}$ or, more precisely,
$S_E \sim (R/R_\text{min}-1)^{1.059^{\pm 0.011}(D-1)-.492^{\pm0.006}}$. 
This result ensures that $S_E$ is only defined for $D\geq 2$ as it should be.

\begin{figure}[htp]
\includegraphics[width=\columnwidth]{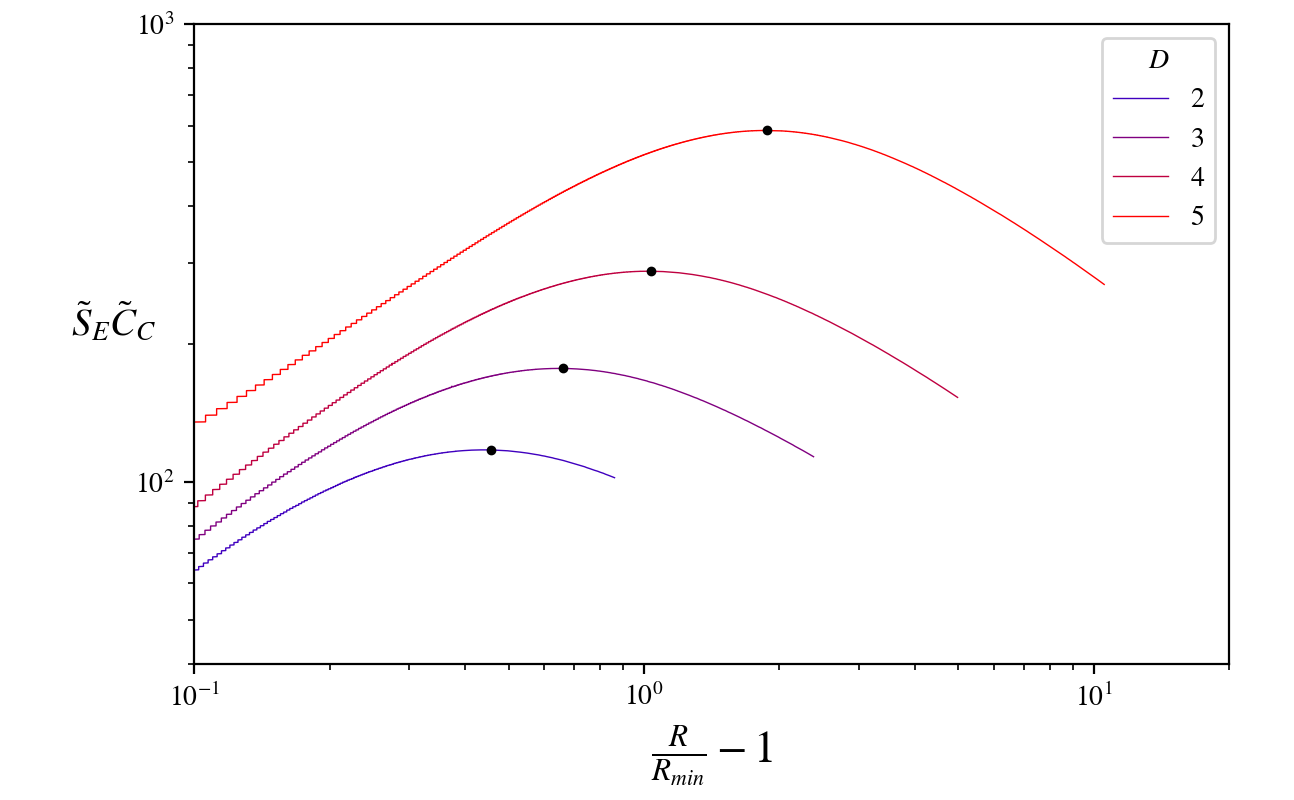}
\caption{The product of the Euclidean Action and the Configurational Complexity versus radius. The maxima, denoted by black dots, discriminate between the thin and thick wall regimes. The near scale invariance of the thin wall regime confirms the numerical estimates of the scaling laws. 
}
\label{SE_times_CC}
\end{figure}

Comparing the results for ${\cal C}_C$ and $S_E$ in the thin-wall limit, we should expect the scaling to go as ${\cal C}_C \sim  S_E^{-1}$, independent of spacetime dimensions. 
This scaling is suggested in Figure \ref{CC_fields_vs_SE}, for large $S_E$. In Figure \ref{SE_times_CC}, we plot the product of $\tilde S_E \tilde {\cal C}_C$ versus radius, showing that the scaling holds quite well in the thin wall limit, with a best fit leading to $\tilde {\cal C}_C \tilde S_E \sim (R/R_\text{min}-1)^{0.052^{\pm 0.005} (D-1) + 0.256^{\pm 0.019}}$, implying a very weak dimensional dependence. Also note the clear change of behavior between thick and thin wall instantons, denoted by the dots in the Figure. As expected, DCC can be used a shape discriminator.

In the thick-wall limit, $S_E/S_E^\text{min} \gtrsim 1$, expanding Eq. \ref{Lorentzian1} we get $\tilde {\cal C}_C \simeq \left [1-b( \tilde S_E -1)^c\right ]$, or, $\tilde{\cal C}_C\sim 1$. 
The trend to a constant value for small $R\sim R_\text{min}$ is clear in Fig. \ref{CC_versus_R}.
Instantons with minimal action are the most unstable and thus have maximum configurational complexity, illustrating the connection between configurational complexity and instability, as observed before in the context of stars \cite{Sowinski:2013}, oscillons \cite{Sowinski:2018b}, and other configurations \cite{Correa:2014}. 
These two limits allow us to estimate the power law in the Lorentzian fit as $c\simeq 1$. 
The best-fit value is found numerically to be $c = 1.002^{\pm .112}-.05^{\pm.0006}(D-1)$, so with a very weak dimensional dependence. 

%
%---------------------------------------------------------------------------------------------SECTION 4
\section{Conclusions and Outlook}\label{s5}
In this paper we have applied the framework of Configurational Information Measures (CIMs) to instantons in a scalar field theory in $D$ spaetime dimensions with an asymmetric double well potential.
We have focused on the differential configurational complexity (DCC) of these profiles in order to understand the informational complexity inherent of their shapes. 
We have a found a clear relation between the stability of the vacuum (related to the Euclidean action of the instanton) and the instanton's DCC.
We've obtained a new power law behavior of DCC in the thin wall limit, and an asymptotic complexity in the thick wall limit, independent of the instanton's action. 
We have also shown that the Euclidean action of instanton solutions is related to the DCC in such a way as to serve as a pattern discriminator between thin and thick wall configurations. 
There is a clear inverse relation between the instanton stability--related to the lifetime of the vacuum--and its corresponding DCC for all $D\leq 5$: longest-lived vacua have instantons with largest Euclidean action and smallest DCC, expressed as a Lorentzian fit. 
Finally, we have found a scaling relation between the Euclidean instanton action and the DCC, which is nearly independent of the spacetime dimensionality. 

Possible extensions of this work would include the application of CIMs to the evolution of bubbles post nucleation, as well as different types instantons. 
The former would help characterize the informational properties of first-order vacuum phase transitions. 
The latter would help elucidate the informational properties of Fubini-Lipatov and Coleman-de Luccia instantons, although we expect similar qualitative behavior to what we have found here.
Extensions to thermal phase transitions should be straightforward.
Relating CIMs to standard thermodynamic entropies is crucial for understanding the role shape information plays in the evolution of systems out of equilibrium. 
It would also be interesting to explore generalizations of CIMs away from Shannon-like measures, something that could be accomplished by using extensions of both Renyi and Tsallis-like measures to field theories.

\acknowledgments
MG was supported in part by a Department of Energy grant DE-SC0010386. DS was supported by the Institute for Cross-Disciplinary Engagement at Dartmouth (ice.dartmouth.edu) through a grant from the John Templeton Foundation.

\appendix
\section{Thin Wall Action}\label{a1}
Here we derive the Euclidean Action for the thin-wall instantons.
In the thin wall approximation we treat the vacua as almost degenerate.
The field profile stays close to the core value for a long time, and then abruptly drops to the false vacuum value.
This allows us to ignore the second term of eq. \ref{Instanton EoM},
\ee{
\frac{d^2\phi}{d\xi^2}\approx \frac{dV}{d\phi}.
}
This equation of motion has a first integral 
\ee{
\frac{1}{2}\left(\frac{d\phi}{d\xi}\right)^2-V(\phi) = 0
}
where the RHS is set by the behavior of the solution as $\xi\rightarrow\infty$ - the gradient vanishes, as does the potential in the false vacuum.
This gives us a nice relationship between differentials
\ee{
\frac{d\phi}{d\xi} = -\sqrt{2V(\phi)}.
}
Note that we chose the negative root because the field decays from the core value to the false vacuum. 
With this in hand we can simplify the Euclidean Action coming from eq. \ref{Dimensionless Action} to
\ee{
S_E = 2\int\! d^D\xi \ V(\phi(\xi)).
}

Now we break up the volume integral over three regions: radii smaller than the instanton wall, radii in the instanton wall, and radii outside.
The latter has a vanishing contribution since the false vacuum has vanishing potential.
Since the field drops sharply in the wall, we define the radius of the instanton, $\bar \xi$, implicitly as where the field value drops to half the core value:
\ee{
\frac{v}{2} = \int_0^{\bar\xi}\! d\xi \sqrt{2 V}
}
Recalling the solid angle of a $D$-sphere, $\Omega_D = \frac{2\pi^{\frac{D}{2}}}{\Gamma(\frac{D}{2})}$, the Euclidean action can now be written as
\ee{
S_E &= 2V(v)\Omega_D\int_0^{\bar\xi}\! \!d\xi \ \xi^{D-1}+2\Omega_D\bar\xi^{D-1}\int_{wall}\!\! d\xi \ V(\phi)\nonumber\\
&=-\frac{1}{D}\epsilon v^4\Omega_D\bar\xi^D+\Omega_D\bar\xi^{D-1}\sigma,
}
where we have introduced the wall energy density
\ee{
\sigma = \int_{wall}\!\! d\xi \ V(\phi)=\int_0^{v}d\phi\sqrt{2 V}
}
The action is an extremum with respect to $\bar \xi$, which relates the instanton radius to the energy density
\ee{
\bar\xi =(D-1) \frac{\sigma}{\epsilon\ v^4}.
}
With this in hand, the dimensionless Euclidean action reads
\ee{
S_E = \frac{(D-1)^{D-1}\Omega_D}{D v^{4(D-1)}}\frac{\sigma^D}{\epsilon^{{D-1}}},
}
from which we find the relationship to asymmetry parameter, $S_E\sim\epsilon^{-d}$.

\bibliography{myBib}
\end{document}